# Phosphorene-AsP Heterostructure as a Potential Excitonic Solar Cell Material - A First Principles Study

M. R. Ashwin Kishore[1,2] and P. Ravindran[1,2,3,4, a)]

[1]Department of Physics, Central University of Tamil Nadu, Thiruvarur, Tamil Nadu, 610101, India
[2]Simulation Center for Atomic and Nanoscale MATerials (SCANMAT), Central University of Tamil Nadu, Thiruvarur, Tamil Nadu, 610101, India
[3]Department of Materials Science, Central University of Tamil Nadu, Thiruvarur, Tamil Nadu, 610101, India
[4]Center of Material Science and Nanotechnology and Department of Chemistry, University of Oslo, Box 1033 Blindern, N-0315 Oslo, Norway

[a)]Corresponding author: raviphy@cutn.ac.in

**Abstract.** Solar energy conversion to produce electricity using photovoltaics is an emerging area in alternative energy research. Herein, we report on the basis of density functional calculations, phosphorene/AsP heterostructure could be a promising material for excitonic solar cells (XSCs). Our HSE06 functional calculations show that the band gap of both phosphorene and AsP fall exactly into the optimum value range according to XSCs requirement. The calculated effective mass of electrons and holes show anisotropic in nature with effective masses along Γ-X direction is lower than the Γ-Y direction and hence the charge transport will be faster along Γ-X direction. The wide energy range of light absorption confirms the potential use of these materials for solar cell applications. Interestingly, phosphorene and AsP monolayer forms a type-II band alignment which will enhance the separation of photogenerated charge carriers and hence the recombination rate will be lower which can further improve its photo-conversion efficiency if one use it in XSCs.

## INTRODUCTION

The fossil reserves are our major energy source and it is depleting at a faster rate. Therefore, we need to develop a clean and renewable alternative for fossil fuels. Photovoltaics is the central to the development of a new clean energy economy. It is therefore highly desirable to fabricate or design new materials to improve the current efficiency further. The advent of graphene triggered an enormous research interest on other 2D materials due to their fascinating properties. 2D materials has been successfully employed for wide variety of applications and some of them are electronics, optoelectronics, catalysis, and energy conversion as well as storage. [1] In addition to some of the best known 2D materials such as $MoS_2$, $WS_2$, and MXene, Phosphorene also received great attention due to its thickness-dependent direct electronic bandgap makes promising for nanophotonics and optoelectronics. [2] To utilize solar radiation more effectively, excitonic solar cells (XSCs) are more promising. In this study, we show that phosphorene and arsenic phosphorus are efficient acceptor and donor materials for higher efficiency XSCs.

## COMPUTATIONAL DETAILS

The total energy calculations were performed using the projector augmented wave (PAW) pseudopotentials with the exchange and correlation in the Perdew−Burke−Ernzerhof (PBE) formalism of DFT as implemented in the Vienna *abinitio* simulation package (VASP). [3] We included van der Waals (vdW) correction proposed by Grimme into the calculations. For accurate band structure calculations, we employed Heyd-Scuseria-Ernzerhof (HSE06) hybrid functional with 25% Hartree-Fock exchange energy. Brillouin zone was sampled with a 5×5×1 Monkhorst-Pack **k**-point mesh for PBE functional and 3x3x1 **k**-mesh for more expensive HSE06 calculations. A cutoff energy of 500 eV for the plane wave basis set was used throughout all the calculations. A large vacuum of 15 Å was used to avoid the spurious interaction. The atoms positions and lattice parameters are fully optimized until the force on each atom is less than 0.01 eV/Å.

## RESULTS AND DISCUSSION

Before investigating the phosphorene/AsP (Pn/AsP) heterostructure, let us analyze the electronic properties of phosphorene and AsP monolayer. Phosphorene has a puckered orthorhombic lattice structure with each phosphorus atom connects to three neighboring atoms (see Figure. 1a). AsP monolayer can be obtained by replacing neighboring P atom with As atom. The calculated lattice parameters and the band gap values of phosphorene and AsP monolayer are tabulated in Table 1 and the values are in good agreement with other reports [4,5]. The calculated band structure and density of states (DOS) of phosphorene and AsP monolayer using GGA-PBE are plotted in Figure.1. As shown in Fig.1, both the phosphorene and AsP monolayer are direct band gap semiconductor since the valence band maximum (VBM) and conduction band minimum (CBM) are at the same Γ point in the Brillouin zone. From the DOS plot, we can see that both the VBM and CBM of phosphorene monolayer are dominated by P-3p states with noticeable contribution from P-3s states. On the other hand, both the VBM and CBM of AsP monolayer are dominated by P-3p and As-4p states with noticeable contribution from P-3s and As-4s states. The equal contribution of P-3p and As-4p states to the band edges is due to the almost same electronegativities of both P and As. The well dispersed band features at the band edges and also the hybridization of P-3p and As-4p orbitals show the presence of covalent bonding in these systems which is also typical for 2D layered materials. It is well known that the PBE functional usually underestimate the band gap values of semiconductors. Hence, we have calculated band structure of Pn and AsP monolayer using HSE06 hybrid functional which has been demonstrated to be more accurate in describing the exchange-correlation energy of electrons in solids. It can be seen from Fig.2 (a and b) that phosphorene and AsP monolayer has a direct band gap with the value of 1.55 eV and 1.53 eV, respectively. It is to be noted that the optimum band gap value for a material to be employed for higher efficiency photovoltaic application in single bandgap solar cell is around 1.52 eV. This indicates that Pn/AsP can be an ideal material for higher efficiency solar cell applications.

**TABLE 1.** Lattice constants and band gaps of phosphorene, AsP, and Pn/AsP

| System | $a$ (Å) | $b$ (Å) | Band gap (eV) | |
|---|---|---|---|---|
| | | | PBE | HSE06 |
| phosphorene | 3.30 | 4.56 | 0.87 | 1.55 |
| AsP | 3.51 | 4.61 | 0.87 | 1.53 |
| Pn/AsP (AB-stacked) | 3.40 | 4.56 | 0.45 | - |

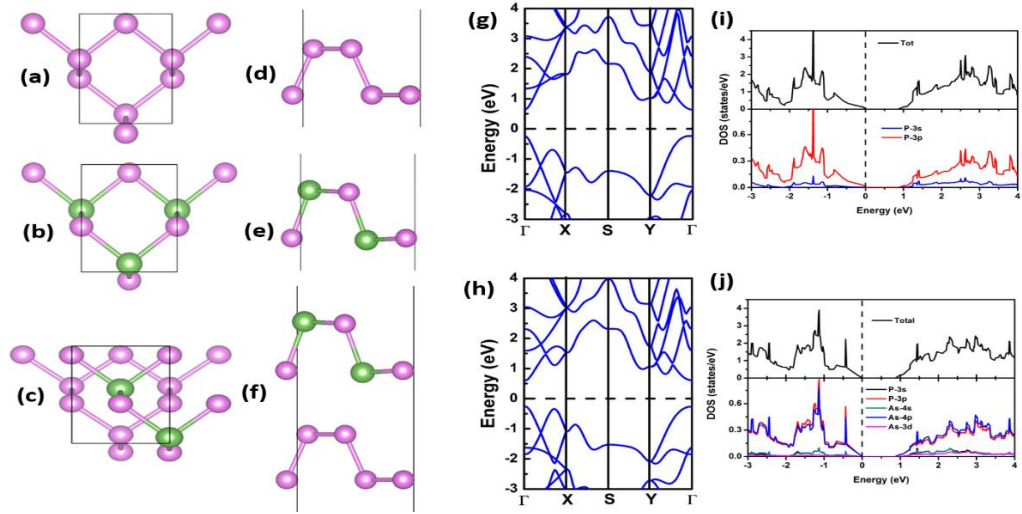

**FIGURE 1.** Top (a,b,c) and side views (d,e,f) of Pn, AsP, and Pn/AsP heterosructure, respectively. Here, the magneta and green balls stand for P and As atoms. (g,h) The calculated band structure and (i,j) corresponding density of states of Pn and AsP monolayers calculated using PBE functional. The Fermi level is set to zero in band structure and DOS plots.

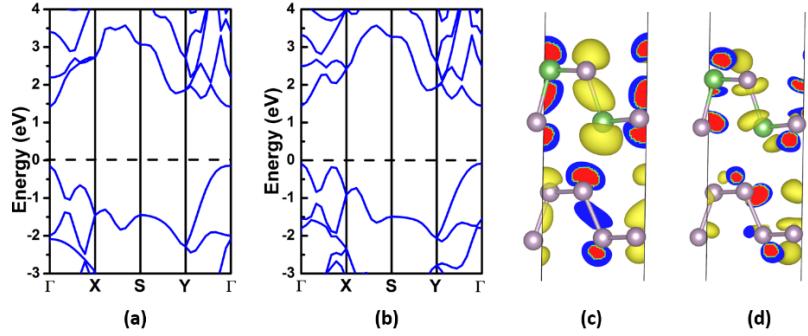

**FIGURE 2.** The calculated band structure of phosphorene (a) and AsP monolayer (b) using HSE06 functional. The Fermi energy is set to zero. (c,d) The isosurface plot of the charge density corresponding to VBM and CBM for AB-stacked Pn-AsP heterostructure, respectively. The isosurface value is 0.004 e Å$^{-3}$.

We now construct the 2D Pn/AsP vdW heterostructure by stacking 1x1 AsP monolayer on 1x1 phosphorene monolayer, and the lattice mismatch is only 1.1% which is good for constructing Pn/AsP heterostructures. Three possible stacking orders are considered in the present study, namely, AA-, AB-, and AC-stacking. Our total energy calculations indicate that the AB-stacking is energetically the most favorable, which is 4 and 8 meV/atom lower than that of AA- and AC-stacking, respectively.

**TABLE 2.** Calculated effective mass of electron/hole for phosphorene and AsP monolayer

| System | Effective mass of electron ($m_0$) | | Effective mass of hole ($m_0$) | |
|---|---|---|---|---|
| | Γ-X | Γ-Y | Γ-X | Γ-Y |
| phosphorene | 0.14 | 1.19 | 0.13 | 7.95 |
| AsP | 0.17 | 1.23 | 0.15 | 1.97 |

The relaxed structure of AB- stacked Pn/AsP heterostructure is shown in Fig.1 (c and f). The interlayer distance between phosphorene and AsP is 3.15 Å which is in the van der Waals range. In Fig.2 (b), we have plotted the isosurfaces of the band decomposed charge density corresponding to VBM and CBM of AB-stacked Pn/AsP heterostructure. One can see that the VBM is contributed from the delocalized states of P and As atoms, while CBM is partially contributed from localized states, especially in the interfacial area between the top and bottom layers. One of the deciding factors in improving the power conversion efficiency in the XSCs is the carrier mobility. The effective mass of charge carrier should be lower than 0.5 $m_o$ at least in one crystallographic direction to have an excellent mobility. [6] Effective mass of both electrons and holes of phosphorene and AsP are listed in Table. 2 It is to be noted that both the effective mass of electrons and holes along the Γ-X direction manifest low effective masses and this can be understood from the band structure (see Fig. 1(g and h)). As one can notice that the VBM and CBM along Γ-X is more dispersed than that of Γ-Y and this leads to low electron/hole effective mass along Γ-X than that of Γ-Y direction.

In addition to an appropriate direct band gap and high carrier mobility, it is important that the material should have good optical absorptivity to enhance the solar conversion efficiency. Optical absorption spectrum of phosphorene and AsP monolayer are calculated and displayed in Figure. 3 (a). We have compensated the band gap underestimation of PBE functional by using a so-called scissor operation with the correct band gap values obtained from HSE06 functional that shift the absorption curves upward in energy. As shown in Figure. 3 (a), the optical absorption for both the phosphorene and AsP monolayer increases significantly with photon energy over the range of visible light and reaches a maximum absorption in ultra violet region. We have observed that the optical absorption coefficient is low even though the band gap of both the materials are in optimum range of 1.5 eV and this can be understood from the DOS. The VBM and CBM of both the materials are mainly originating from p states with little contribution from s states. Since the optical interband transitions obey selection rule, the optical absorption will occurs only when interband optical transitions from s to p state is possible which is minimal in the present cases. Overall, the optical absorption is in a relatively wide energy range from 1.4 eV to 5 eV and this confirms the potential use of these materials as donor and acceptor in higher efficiency XSCs solar cell applications.

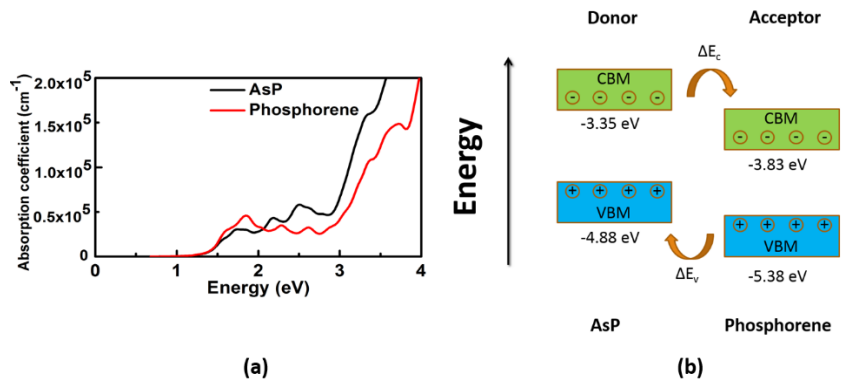

**FIGURE 3.** (a) The calculated optical absorption spectrum of phosphorene and AsP monolayer. (b) The calculated band offsets between phosphorene and AsP monolayer. The numbers are the VBM and CBM levels with respect to vacuum level.

To find the band alignment between the phosphorene and AsP monolayer, we have calculated their band edges using the vacuum level as a common energy reference as shown in Figure. 3 (b). This band alignment further demonstrates the type-II nature of Pn/AsP heterostructure. It is to be noted that the band offsets are crucial factors in the application of XSCs. Thus, we calculated the band offsets $\Delta_{Ec}$ and $\Delta_{Ev}$ of the isolated layers (natural band offsets). The natural band offsets are calculated to be $\Delta_{Ec}$ = 0.48 eV and $\Delta_{Ev}$ = 0.49 eV for Pn/AsP heterostructure. These results confirm that AsP monolayer and phosphorene can potentially serve as efficient donor and acceptor materials in XSCs, respectively.

## CONCLUSIONS

In conclusions, our HSE06 functional calculations show that the band gap of both phosphorene and AsP monolayer fall exactly into the optimum value range according to XSCs requirement. We observed that the effective mass of electrons and holes along Γ-X direction is lower than the Γ-Y direction. Hence the carrier mobility would be higher along Γ-X direction. The wide energy range of light absorption confirms the potential use of these materials for higher efficiency solar cell applications. Interestingly, phosphorene and AsP forms a type-II band alignment which will enhance the separation of photogenerated charge carriers and hence the recombination rate will be lower which can further improve the efficiency of a XSCs. Our overall results confirm that AsP monolayer and phosphorene can potentially serve as donor and acceptor materials, respectively in XSCs.

## ACKNOWLEDGMENTS

The authors are grateful to the Research Council of Norway for computing time on the Norwegian supercomputer facilities. This research was supported by the Indo-Norwegian Cooperative Program (INCP) via UGC Grant No. F.No.58-12/2014(IC) and Department of Science and Technology, India via Grant No. SR/NM/NS-1123/2013.